
\input phyzzx
\PHYSREV
\sequentialequations
\def\as{\alpha_s}
\def\A{{\rm\scriptscriptstyle A}}
\def\T{{\rm\scriptscriptstyle T}}
\def\F{{\rm\scriptscriptstyle F}}
\def\cf{C_\F}
\def\ma{M_\A}
\def\frac#1#2{{#1 \over #2}}

\def\GeV{{\rm GeV}}
\def\ra{\rightarrow}
\def\a{\alpha}

\def\la{\lambda}
\def\g{\gamma}
\def\G{\Gamma}
\def\d{\delta}
\def\e{\epsilon}
\def\s{\sigma}
\def\bg{\bar g}
\def\o{\over}
\def\so{\s_0}
\def\soe{\s_{(\e)}}
\def\l{\left}
\def\r{\right}
\def\ot{\Omega_\T}
{\baselineskip 14pt
\null
\rightline{BNL-49061}
\vskip 1.5cm
\centerline{\bf QCD CORRECTIONS TO PRODUCTION OF HIGGS PSEUDOSCALARS}
\vskip 1.5cm
\centerline{{\bf  Russel P.~Kauffman and Wendy Schaffer}
\foot{This manuscript has been authored
under contract number DE-AC02-76 CH 00016 with the U.S. Department of Energy.
Accordingly, the U.S. Government retains a non-exclusive, royalty-free
license to publish or reproduce the published form of this contribution, or
allow others to do so, for U.S. Government purposes.}}

\vskip 1.cm
\centerline{{\it Physics Department}}
\centerline{{\it Brookhaven National Laboratory, Upton, NY~~11973}
\foot{Internet: kauffman@bnlcl1.bnl.gov, wendys@bnlcl1.bnl.gov}}
\vskip 1.5cm
\centerline{ABSTRACT}
\vskip 0.5cm

Models of electroweak symmetry breaking with more than a single
doublet of Higgs scalars contain a neutral pseudoscalar boson.
The production of such a pseudoscalar in hadron collisions proceeds
primarily via gluon fusion through a top-quark loop (except
for those models in which the pseudoscalar coupling to bottom
quarks is strongly enhanced).
We compute the QCD corrections to this process in the heavy-quark limit,
using an effective Lagrangian derived from the axial anomaly.
\vskip 0.5cm

\vfill
\line{April 1993\hfill} }
\vfill
\eject
The primary mission of the next generation of colliders is
the study of the physics of electroweak symmetry breaking.
Since the Higgs mechanism provides
the most promising scenario for the symmetry breaking,
the search for the particles of the Higgs sector is of primary importance.
The minimal model of electroweak symmetry breaking contains one complex
scalar doublet, three components of which become the longitudinal degrees
of freedom of the $W^{\pm}$ and $Z$.  The remaining component of the doublet
is the so-called Higgs boson.  However, the minimal model with one doublet
has no {\it a priori} justification and there are several motivations
for considering models with enlarged Higgs sectors, either containing
more doublets, singlets or more exotic representations.  For example,
supersymmetric models require at least two doublets.  Likewise, at least
\REF\guide{J.F.~Gunion, H.E.~Haber, G.L.~Kane, and S.~Dawson, {\it The Higgs
Hunter's Guide} (Addison-Wesley, Menlo Park, California, 1990), p. 201.}
two doublets are required to produce CP violation in the Higgs sector[\guide].

Models with enlarged Higgs sectors have a richer particle content than
the minimal model; in general, neutral pseudoscalars (with respect to
their fermion couplings)
and charged scalars
as well as extra neutral scalars are present. In this letter we will study
the QCD corrections to the production of a Higgs pseudoscalar ($A^0$) via
gluon fusion. This process proceeds primarily through a top-quark loop
(unless the coupling to bottom quarks is greatly enhanced).
We will focus on the case of a light pseudoscalar and work in the
heavy-top-quark limit : $m_t \gg \ma$.

\REF\model{In a generic two-Higgs-doublet model[\guide] there is also a factor
of $v_1/v_2$ with $v_1~(v_2)$
being the vacuum expectation value of the Higgs doublet to which the top
couples (does not couple).}
To fix our normalization we take
the coupling of the pseudoscalar to
top quarks to be $m_t \gamma_5 /v$ where $v=246~{\rm GeV}$[\model].
The lowest-order amplitude for $ gg\ra A^0$ is well known and takes the form
\REF\agg{J.F.~Gunion, G.~Gamberini, and S.F.~Novaes, Phys. Rev. D {\bf 38},
3481 (1988).}
[\agg]
$$
{\cal M}(g(k_{1\mu}^a),g(k_{2\nu}^b)\ra A^0) = -i g_A \delta^{ab}
        \e^{\mu\nu\rho\sigma} k_{1\rho} k_{2\sigma} \tau f(\tau),
\eqn\lowest$$
with $\tau=4m_t^2/\ma^2$, $g_A =  \as/2\pi v$, and
$$
f(\tau) = \cases{ \l[\sin^{-1}\l(\sqrt{1/\tau}\r)\r]^2,
          & if $\tau\geq 1$, \cr
          -1/4 \l[\ln(\eta_+/\eta_-) - i\pi \r]^2,
          & if $\tau < 1$, \cr }
$$
where $\eta_\pm = (1 \pm \sqrt{1-\tau})$.
We see that in the large $m_t$ limit the
amplitude for $gg \ra A^0$ is
independent of the top quark mass, just as in the scalar case:
$\tau f(\tau)\ra 1$ as $\tau\ra\infty$.
The heavy-top approximation of the amplitude is accurate to within 5\%
for $m_t^2  > 2\ma^2$ and to within 10\% for $m_t^2 > \ma^2$.

The amplitude in Eq. \lowest\  can be computed by evaluating the
triangle diagram with a top quark in the loop and taking the limit
$ m_t\ra\infty $ or instead by
\REF\abj{S.L.~Adler, Phys. Rev. {\bf 177}, 2426 (1969);
         J.~Bell and R.~Jackiw, Nuovo Cim. {\bf 60A}, 47 (1969).}
noticing that this amplitude is related to the axial anomaly[\abj].  Let
$j_5=\bar\psi \g_5\psi$ be the axial current and
$j_5^\mu=\bar\psi i\g_\mu\g_5\psi$ be the axial vector current.
The anomaly equation reads:
$$ \partial_\mu j_5^\mu = -2 m_t j_5
           +i{\as \over 8\pi} G_{\mu\nu}^a \tilde{G}_{\mu\nu}^a,
\eqn\anomaly$$
where $G_{\mu\nu}^a$ is the field-strength tensor for ${\rm SU}(3)$
and $\tilde{G}_{\mu\nu}^a$ is its dual,
$\tilde{G}_{\mu\nu}^a = \e^{\mu\nu\rho\sigma} G_{\rho\sigma}^a$.
In the heavy quark limit the left side of Eq. \anomaly\
vanishes; it is proportional
\REF\suth{
D.~Sutherland, Nucl. Phys. {\bf B2}, 433 (1967).
}
to more powers of external momenta than the other terms[\suth].  Therefore,
in the heavy quark limit the matrix element of $j_5$ between gluon
states is given
$$ 2 m_t \langle g | j_5 | g \rangle =
  {i\as \over 8\pi} \langle g | G_{\mu\nu}^a \tilde{G}_{\mu\nu}^a |g \rangle.
\eqn\matrix$$
Equivalently, one may treat the interactions of gluons with the $A^0$
\REF\sally{Effective Lagrangians have been used to describe the interactions
of a scalar Higgs boson with gluons. See S.~Dawson,
Nucl. Phys. {\bf B359}, 283 (1991); A.~Djouadi, M.~Spira and
P.M.~Zerwas, Phys. Lett. {\bf B264}, 440 (1991); and references therein.}
in the heavy-quark limit as arising from the effective Lagrangian[\sally]
$$ {\cal L}_{\rm eff} = {\as \over 8\pi v}
                      G_{\mu\nu}^a \tilde{G}_{\mu\nu}^a A^0.
\eqn\leff $$
The power of Eq. \leff\ for our problem comes from the Adler-Bardeen theorem
\REF\aberdeen{S.L.~Adler and W.A.~Bardeen, Phys. Rev. {\bf 182}, 1517 (1969).
 For explicit proofs to second order in the non-abelian theory see
              R.~Akhoury and S.~Titard, Report No. UM-TH-91-21 (unpublished);
              M. Bos, Report No. UCLA/92/TEP/41 (unpublished).}
[\aberdeen]
which states that Eq. \anomaly\ is true
{\it to all orders in perturbation theory.}
Therefore, Eq. \leff\ provides the correct effective Lagrangian with which to
compute radiative corrections.  The advantage is that amplitudes which
correspond to two-loop diagrams in the original theory are one-loop diagrams
in the effective theory. A similar Lagrangian can be written for the
\REF\zerwas{A.~Djouadi, M.~Spira and P.M.~Zerwas,
Report No. DESY 92-170 (to be published).}
\REF\hphotons{In the heavy top-quark limit QCD corrections to scalar Higgs
decay to two photons may be calculated using an effective Lagrangian
(see Ref. [\sally]) or
directly through the full two-loop diagrams: H.~Zheng and D.~Wu,
Phys. Rev. D {\bf 42}, 3760, (1990); A.~Djouadi, M.~Spira, J.~van der Bij
and P.M.~Zerwas, Phys. Lett. {\bf B257}, 187 (1991); S.~Dawson and
R.P.~Kauffman, Phys. Rev. D {\bf 47}, 1264 (1993).}
interaction of photons with the pseudoscalar[\zerwas,\hphotons];
in that case one sees
immediately that the ${\cal O}(\as)$ corrections to $\g\g\ra A^0$ vanish
in the $m_t\ra\infty$ limit.

The effective Lagrangian, Eq. \leff, leads to two- and three-gluon
vertices with the $A^0$ (as well as a four-gluon vertex which is
irrelevant to the current study):
$$\eqalign{
V_3(k_{1\mu}^a,k_{2\nu}^b) & = -i g_A \delta^{ab} \e^{\mu\nu\rho\sigma}
                               k_{1\rho} k_{2\sigma}, \cr
V_4(k_{1\mu}^a,k_{2\nu}^b,k_{3\rho}^c) & = g g_A f^{abc} \e^{\mu\nu\rho\sigma}
                               (k_1 + k_2 + k_3)_\sigma, \cr
}\eqn\vert$$
where the $k_i$ are the gluon momenta directed inward.

We use dimensional regularization in computing the radiative
corrections. However, the $\e$-tensors in Eq. \vert\
are intrinsically four-dimensional
objects and must be treated as such.  The product of two $\e$-tensors
can be written in terms of $\bg^{\mu\nu}$ the metric tensors in the
four-dimensional sub-space:
$$
-\e^{\a\la\mu\nu} \e_{\a\rho\s\tau}
  = \bg^\la_\rho(\bg^\mu_\s\bg^\nu_\tau - \bg^\mu_\tau\bg^\nu_\s )
   - \bg^\la_\s(\bg^\mu_\rho\bg^\nu_\tau - \bg^\mu_\tau\bg^\nu_\rho )
   + \bg^\la_\tau(\bg^\mu_\rho\bg^\nu_\s - \bg^\mu_\s\bg^\nu_\rho ),
\eqn\epsq$$
where $\bg^{\mu \nu} \bg{\mu \nu} = 4$.
For simplicity, we take the incoming momenta to be in four dimensions
(this is simply a choice of frame).
The lowest order cross section is then, averaged over colors and
polarizations in $n=4-2\e$ dimensions,
$$\eqalign{
\sigma^{(0)}(g(k_1)g(k_2)\ra A^0)
      &= {\pi g_A^2 \over 8 (N_c^2-1)(1-\e)^2}\delta(1-z)\cr
      &\equiv \soe\delta(1-z)
       \equiv {1\over(1-\e)^2}\so\delta(1-z),\cr
}\eqn\sigoe$$
where $z=\ma^2/s$ and $s = (k_1+k_2)^2 $.

\FIG\figone{Feynman diagrams for the virtual corrections to $gg\ra A^0$.}
The virtual radiative corrections are given by the diagrams in Figure 1,
where the amplitude involving the four-gluon vertex vanishes due to the
antisymmetry of the $ggA^0$-vertex. The other diagram yields the result
for the virtual corrections
$$\s_v = {N_c\as\soe\o\pi}\eta
             \l( -{1\o\e^2} + {2\o3}\pi^2 + 2 \r)\d(1-z), \eqn\virtual $$
where
$$ \eta = \G(1+\e)\l({4\pi\mu^2\o\ma^2}\r)^\e. $$

\FIG\figtwo{Feynman diagrams for the processes a) $gg\ra gA^0$ and
 b) $q\bar q\ra gA^0$.  The diagram for $q g\ra qA^0$ is a crossing
of diagram b.}
The real corrections are given by the diagrams in Figure 2.
The amplitudes squared for the various processes are, averaged
over colors and spins,
$$\eqalign{
|{\cal M}(gg\ra gA^0)|^2 &= 8 N_c \as\soe
      \l\{ {s^4{+}t^4{+}u^4{+}\ma^8\o stu} - 2\e s\l({u\o t} {+} {t\o u}\r)
      + 2 \hat k_3\cdot\hat k_3 s^2 \l({1\o t^2}+{1\o u^2}\r) \r\},
\cr
|{\cal M}(qg\ra qA^0)|^2 &= 8 \cf\as\soe(1-\e)
      \l\{ -{s^2+u^2\o t} + 2{s^2\o t^2} \hat k_3\cdot\hat k_3 \r\},
\cr
|{\cal M}(q\bar q\ra gA^0)|^2 &= 8 \cf\as\so
     {(t^2+u^2)\o s},
\cr}\eqn\realm$$
where $s$, $t$, and $u$ are the familiar Mandelstam variables and
$k_3$ is the momentum of the final state quark or gluon, with the
hat denoting the $(n-4)$-dimensional components. Since the initial
state particles are taken to be in four dimensions $\hat k_3\cdot\hat k_3$
is the only quantity which depends on the $(n-4)$-dimensional components.
Under integration over the angular variables transverse to the incoming
particles we find
\REF\hatint{J.G.~K\"orner, G. Schuler, G. Kramer, and B. Lampe,
Phys. Lett. {\bf 164B}, 136 (1985).}
[\hatint]
$$\int d\ot \hat k_3\cdot\hat k_3 = -{(n-4)\o(n-2)}k_\T^2
  \int d\ot,\eqn\khat $$
where $k_\T$ is the transverse momentum of the outgoing gluon.
So the terms with $\hat k_3\cdot\hat k_3$ will contribute terms of
${\cal O}(1)$, but only when multiplied by double poles, $t^{-2}$ or $u^{-2}$
(this was used in simplifying Eq. \realm).

Integration over $n$-dimensional phase space gives the cross sections
$$\eqalign{
\s(gg\ra gA^0) &= {\as N_c\soe\o \pi}\eta^\prime
              \bigg\{ \d(1-z)\l({1\o\e^2}-{\pi^2\o3}\r)
              -{2z\o\e}\l[{z\o(1-z)_{+}} + {1-z\o z} + z (1-z) \r]
\cr
   &\qquad\qquad\qquad + 2\l({\log(1-z)\o1-z}\r)_{+} \l[1+(1-z)^4+z^4\r]
             -{11\o6}(1-z)^3   \bigg\},
\cr
\s(qg\ra qA^0) &= {\as \cf\soe\o 2\pi}(1-\e)\eta^\prime
              \bigg\{ \l(-{1\o\e}+2\log(1-z)\r)\l[1+(1-z)^2\r]\cr
      &\qquad\qquad\qquad\qquad\qquad  -{1\o2}(1-z)(7-3z) \bigg\},
\cr
\s(q\bar q\ra gA^0) &= {2\as \cf^2\so\o 3\pi}(1-z)^3,
\cr }\eqn\reals$$
where
$$ \eta^\prime = \G(1+\e)\l({4\pi\mu^2\o s}\r)^\e. $$
It is interesting to note that the real corrections, when written in terms
of the lowest-order cross section are identical in form to those
for the scalar case[\sally].

We see that the terms of ${\cal O}(1/\e^2)$ cancel between the real
and virtual diagrams.  The terms of ${\cal O}(1/\e)$ may be absorbed
into redefinitions of the parton distribution functions in the usual
factorization procedure. We use the $\overline{\rm MS}$ prescription
which fixes the subtraction terms as
$$\eqalign{
\s_{\rm AP}^{gg} &= {\as\o 2\pi}\soe 2 zP_{gg}(z)(4\pi)^\e\G(1+\e){1\o\e},\cr
\s_{\rm AP}^{qg} &= {\as\o 2\pi}\soe zP_{gq}(z)(4\pi)^\e\G(1+\e){1\o\e},\cr
}\eqn\sap$$
where
the splitting functions are defined
$$\eqalign{
P_{gg}(z) &= 2N_c\l[{z\o(1-z)_{+}} + {1-z\o z} + z (1-z) \r]
            + \beta_0 \delta(1-z) \cr
P_{gq}(z) &= \cf{\l[1+(1-z)^2\r] \o z}, \cr
}\eqn\split$$
where $\beta_0 = {11\o6}N_c - {1\o3}N_f$ with $N_f$ being the number
of light quark flavors. The charge renormalization subtraction term is
$$
\s_{\rm ch} = - {\as\o\pi}(4\pi)^\e\G(1+\e)\soe \d(1-z) {\beta_0\o\e}.
\eqn\charge$$

When all the contributions are included the final results are
$$\eqalign{
\hat\s_{gg} &= {\as N_c \so\o\pi}\bigg\{
           \d(1-z)\l({\pi^2\o 3} + 2 \r)
           + 2 z\ln\l({\ma^2\o z\mu^2}\r)
                  \l[{z\o(1-z)_{+}} + {1-z\o z} + z (1-z) \r] \cr
    &\qquad\qquad + 2 \l({\ln(1-z)\o 1-z}\r)_+
           \l[1+(1-z)^4+z^4\r] - {11\o 6}(1-z)^3 \bigg\},
\cr
\hat\s_{qg} &= {\as \cf \so\o 2\pi}\bigg\{
             \l[1 + \ln\l({\ma^2(1-z)^2 \o z\mu^2 }\r)\r]\l[1+(1-z)^2\r] \cr
   & \qquad\qquad\qquad\qquad\qquad - {1\o2}(1-z)(7-3z) \bigg\}.
\cr}\eqn\final $$
Since the real corrections in our result are the same as for the scalar case,
the only difference between the scalar and pseudoscalar results
is the coefficient of the
$\d(1-z)$ term coming from the virtual result:  $N_c(\pi^2/3 + 2)$ for the
pseudoscalar and $(N_c\pi^2/3 + 5/2N_c - 3/2\cf)$ for the scalar.
The numerical similarity of the two constant terms [an accident of SU(3)]
means that the ratio between the
next-to-leading order result and the lowest-order result, or the `K-factor,'
will be almost equal for the two processes.

\REF\hmrs{
P.~Harriman, A.~Martin, W.J.~Stirling, and R.~Roberts, Phys. Rev. D {\bf 42},
798 (1990).}
\FIG\figthree{Radiatively corrected results ($\s_{\scriptscriptstyle TOT}$)
for $A^0$ production in
proton-proton collisions at the SSC, $\sqrt{S} = 40~{\rm TeV}$, with parton
distributions from Ref. \hmrs\ for several values of the
factorization/renormalization scale $\mu$. Shown for comparison is the
lowest order cross section (computed with the two-loop $\as$) with $\mu=\ma$.}
\FIG\figfour{Same as Figure 3 for the LHC, $\sqrt{S} = 15.4~{\rm TeV}$.}
The radiatively corrected cross section is plotted for proton-proton
collisions at $\sqrt{S} = 40~{\rm TeV}$
and $\sqrt{S} = 15.4~{\rm TeV}$
in Figure 3 and Figure 4 for several choices of renormalization
scales and using HMRSB parton distributions[\hmrs].
The ratio between the radiatively corrected result and the lowest order result
(computed with the two-loop $\as$)
ranges from about 2.6 for $\ma = 50~\GeV$
to about 2.2 for $\ma = 200~\GeV$ for both values of $\sqrt{S}$.
As is the case for the scalar Higgs, the contribution from the $gg$ initial
state dominates with about half the correction coming from the $\d(1-z)$ term.
\ack
The authors thank R. Akhoury, S. Dawson, W. Marciano and A. Stange
for useful discussions.
\endpage
\refout
\endpage
\figout
\bye